
\font\twelverm=cmr12			\font\twelvei=cmmi12
\font\twelvesy=cmsy10 scaled 1200	\font\twelveex=cmex10 scaled 1200
\font\twelvebf=cmbx12			\font\twelvesl=cmsl12
			\font\twelveit=cmti12
	
\skewchar\twelvei='177			\skewchar\twelvesy='60

\def\rm{\fam0\twelverm}          \def\it{\fam\itfam\twelveit}%
  \def\sl{\fam\slfam\twelvesl}     \def\bf{\fam\bffam\twelvebf}%
                   \def\cal{\fam 2}%

  \textfont0=\twelverm   \scriptfont0=\tenrm   \scriptscriptfont0=\sevenrm
  \textfont1=\twelvei    \scriptfont1=\teni    \scriptscriptfont1=\seveni
  \textfont2=\twelvesy   \scriptfont2=\tensy   \scriptscriptfont2=\sevensy
  \textfont3=\twelveex   \scriptfont3=\twelveex  \scriptscriptfont3=\twelveex
  \textfont\itfam=\twelveit
  \textfont\slfam=\twelvesl
  \textfont\bffam=\twelvebf \scriptfont\bffam=\tenbf
  \scriptscriptfont\bffam=\sevenbf
\newcount\numco\numco=0
\def\nono#1$${#1\eqno(\the\numco)$$}
\everydisplay{\global\advance\numco by 1\nono}
\def\cite#1{#1}
\headline={\ifnum\pageno=1\firstheadline\else
\ifodd\pageno\rightheadline \else\leftheadline\fi\fi}
\def\firstheadline{\hfil}
\def\rightheadline{\hfil}
\def\leftheadline{\hfil}
        \footline={\ifnum\pageno=1\firstfootline\else\otherfootline\fi}
\def\firstfootline{\rm\hss\folio\hss}
\def\otherfootline{\hfil}
\font\tenbf=cmbx10
\font\tenrm=cmr10
\font\tenit=cmti10
\font\elevenbf=cmbx10 scaled\magstep 1
\font\elevenrm=cmr10 scaled\magstep 1
\font\elevenit=cmti10 scaled\magstep 1

\font\ninerm=cmr9

\font\sevenrm=cmr7
\def\part{\partial}
\def\al{\alpha}
\def\del{\delta}
\def\be{\beta}
\def\gam{\gamma}

\def\l{\ell}

\def\lam{\lambda}

\def\ie{{\it i.e.,}\ }

\def\ee{\hbox{e}}
\def\dd{\hbox{d}}
\def\dD{\hbox{D}}
\def\ttt{{2\over3}}
\def\ths{{3\over2}}
\def\vw{\hbox{vol.(Weyl)}}

\def\hg{{\hat g}}

\def\ssc{\scriptscriptstyle}
\def\pref#1{${}^{#1}$}
\def\cite#1{[#1]}
\def\papers{2}
\def\pol{1}
\def\david{3}

\nopagenumbers
\line{\hfil }
\vglue 1cm
\hsize=6.0truein
\vsize=8.5truein
\baselineskip=13truept
\rightline{\elevenrm iassns-hep-92-36; McGill/93-44}
\rightline{\elevenrm hep-th/9311083}
\vskip 1truecm
\parindent=3pc
\baselineskip=10pt
\font\tif=cmr10 scaled\magstep3
\centerline{\tif Conformally Invariant Off-shell %
Strings\footnote{${}^*$}{\ninerm %
\baselineskip=11pt Talk by R.C.M. at Strings '93, Lawrence Berkeley %
Laboratory}}
\vglue 1.0cm
\centerline{\tenrm ROBERT C. MYERS}
\baselineskip=13pt
\centerline{\tenit Physics Department, McGill University, Ernest
Rutherford Building,}
\baselineskip=12pt
\centerline{\tenit Montr\'eal, Qu\'ebec, H3A 2T8, Canada}
\vglue 0.3cm
\centerline{\tenrm and}
\vglue 0.3cm
\centerline{\tenrm VIPUL PERIWAL}
\centerline{\tenit The Institute for Advance Study}
\baselineskip=12pt
\centerline{\tenit Princeton, New Jersey, 08540-4920, U.S.A.}
\vglue 0.8cm
\centerline{\tenrm ABSTRACT}
\vglue 0.3cm
{\rightskip=3pc
 \leftskip=3pc
 \tenrm\baselineskip=12pt
 \noindent
Recent advances in non-critical string theory
allow a unique continuation of critical Polyakov string amplitudes to
off-shell momenta, while preserving conformal invariance.
These continuations
possess unusual, apparently stringy, characteristics, as we illustrate
with our results for three-point functions.
\vglue 0.8cm }
\line{\elevenbf 1. The Polyakov functional integral\hfil}
\vglue 0.4cm
\baselineskip=14pt
\elevenrm
The Polyakov path integral\pref{\pol} appears only a recipe for perturbative
computations of scattering amplitudes, yet it serves as the foundation
for our entire understanding of string theory. It is natural to ask whether
this framework might yield any insight into the off-shell
properties of strings.
We report on how the recent progress in understanding non-critical
strings provides a new technique to calculate off-shell Polyakov
amplitudes\pref{\papers}. This approach preserves conformal invariance,
and so these amplitudes lend themselves to analysis
by essentially the standard techniques of conformal field theory.

Space-time scattering amplitudes of string excitations are calculated
as correlation functions of vertex operators in a functional integral
over the metric on the string world-sheet and the space-time
string configurations:
$$
\bigg\langle \prod_i \int\!\dd^2\!z_i \sqrt g \,V_i(z_i)\bigg\rangle
\equiv
\int {\dD g\ \dD X\over \hbox{vol.}(\hbox{Diff}\times
\hbox{Weyl})}\ \ee^{-S[g,X]}
\ \prod_i \int\! \dd^2\!z_i \sqrt g\, V_i(z_i)\ .
$$
The measure is divided by the `volume' of the symmetries of the
classical action $S\equiv 1/(8\pi)
\int\!\dd^2\!z \sqrt g g^{ab}\part_a X^\mu\part_b
X_\mu,$ with $\mu=1,\dots,D$ --- namely, diffeomorphisms and local
Weyl rescalings on the world-sheet. Choosing conformal gauge, $g_{ab}
\equiv\ee^{2\phi}\hg_{ab}(m)$, and fixing diffeomorphisms \`a la
Faddeev-Popov, one finds that these functional integrals reduce to
\def\ordre{2}
$$
\int\!\dd m~{\dD \phi\over\hbox{vol.}(\hbox{Weyl})}
{\dD X\ \ \hbox{Det}_{\ssc\rm FP}' \over\hbox{vol.(c.k.v.)}}
\ \ee^{-S[\hg,X]}
\ \prod_i \int\! \dd^2\!z_i \sqrt {\hat g(m)} V_i(z_i)\ \ .
$$
Here,
c.k.v.~stands for the conformal Killing vectors which arise
if the world-sheet is a sphere or a torus,
and $\dd m$ denotes the measure for
integrating over the moduli of surfaces with one or more handles.
In Eq.~(\ordre), the Weyl factor should decouple from the theory,
and the integration $\dD\phi$ would
cancel against the volume of the
group of Weyl rescalings in the denominator.
This decoupling is only actually achieved if
Weyl rescaling survives as a symmetry of the quantum path integral.
This requires\pref{\pol} that $D=26$ in order to cancel
the anomalous dependences on the Weyl field in the
measure factor,
$\dD X\,\hbox{Det}_{\ssc\rm FP}'/\hbox{vol.(c.k.v.)}$.
Also, one must impose various {\elevenit space-time}
 mass-shell and polarization/gauge
conditions on the external string states to avoid any anomalous Weyl
dependences from normal-ordering the vertex operators. Combined, these
restrictions ensure that $\phi$ is decoupled
from {\elevenit on-shell} correlation functions in {\elevenit critical}
string theory,
and the Weyl factor simply disappears from the functional integral.

\def\dtx{\!\dd^2\!x}
\def\dtz{\!\dd^2\!z}
Therefore the mass-shell conditions are obtained from requiring Weyl
invariance. It follows, in the Polyakov approach, that the
calculation of amplitudes for {\elevenit off-shell}
string states requires the ability to compute correlation functions
of vertex operators with an anomalous Weyl dependence,
in the normalized measure $\dD\phi/\vw .$
Why are such computations difficult? The problem resides in the
non-linearity of the Riemannian metric that defines $\dD\phi.$
The norm on infinitesimal changes of the conformal factor is constructed
with the full world-sheet metric $g_{ab}$
$$
(\delta\phi,\delta\phi) = \int\dtx\sqrt{g}(\delta\phi)^2=
\int\dtx \sqrt{\hat g}\,\ee^{2\phi}(\delta\phi)^2,
$$
which then explicitly depends on $\phi.$  The functional integral over
$\phi$ would be a standard quantum field theory with
the measure, $\dD_{\ssc 0}\phi,$
defined by the translation-invariant norm,
$(\delta\phi,\delta\phi)_{\ssc 0} = \int\dtx \sqrt{\hat g}
(\delta\phi)^2.$
The relation between these two measures is remarkably simple,
\def\mesure{4}
$$\dD\phi = \dD_{\ssc 0}\phi\ \exp\left(S_{\ssc L} -
{\mu\over \pi}\int \dtz \,\ee^{\al\phi}\right),
$$
where
$S_{\ssc L} \equiv \int {\,\dtz\over6\pi} \bigg[\part\phi\bar\part\phi
+ {1\over4} \sqrt {\hat g} \hat R\phi \bigg]$.
The `cosmological constant' $\mu$ is the coefficient of a local
counterterm, and remains undetermined in this computation.
The constant $\al$ in this interaction is explicitly fixed (see below).
This relation (\mesure) originally arose
in the study of two-dimensional gravity coupled
to conformal matter\pref{\david}.
It is important to note that
the derivation of Eq.~(\mesure) is entirely
independent of the rest of the functional integrals involved.
Thus it remains valid for our studies of
off-shell amplitudes.
\vglue 0.6cm
\line{\elevenbf 2. Correlation functions\hfil}
\vglue 0.4cm
To proceed further, we assume that the correlation functions of interest
may be calculated with conformal field theory methods.
For non-critical strings, this approach has been
verified by comparison with results determined
by matrix model techniques.
The stress tensor deduced from $S_{\ssc L}$ is
$T_{\ssc L} = {1\over 6} \left[(\part\phi)^2-\part^2\!\phi\right]$,
which makes no contribution to the total central charge.
Off-shell vertex operators receive exponential Weyl dressings,
$\ee^{\be\phi}V$ (just
as matter operators in non-critical string theory do), where
$$\be={1\over 6}\left[\sqrt{1-12\del}-1\right].$$
Here, $\del=k^2+2n$ (with $n=-1,0,1,\ldots$) is the mass-shell operator
of the particular
external string state. Explicitly, off-shell tachyon vertex
operators are $\ee^{\be\phi}\ee^{ik\cdot X}$ with $\del=k^2-2$, and
hence $\be=(\sqrt{25-12k^2}-1)/6.$ The dressing
exponents are chosen to vanish when the vertex operator goes
on-shell (\ie $\del=0$), which insures that
the off-shell amplitudes reduce precisely to
the usual on-shell amplitudes.
Rather puzzling is the non-analyticity in this prescription
at $\del={1\over 12}.$ Further insight into quantum Liouville
theory is needed to overcome this barrier, but
for the present time, we will limit our attention to $\del\le {1\over 12}$
or $\be\ge-{1\over6}$.
Treating the cosmological constant term within the conformal
field theory context fixes $\al= \ttt$ in Eq.~(4).

Explicit calculations may be carried out for tree-level scattering
amplitudes. We omit the details here, but as an example present the
result for an off-shell tachyon three point amplitude
${\cal A}=\langle\prod_{i=1}^3\int\!\dd^2\!z_i\ \ee^{\beta_i\phi}
\,\ee^{ik_i\cdot X}\rangle$.
The final result, valid for arbitrary external momenta subject only
to the constraint $\del_i\le{1\over12}$ or $k_i^2\le{25\over12}$, is
$$
{\cal A}=X\
\Gamma(\hbox{${1+3\gam\over2}$})\Gamma(\hbox{${1-3\gam\over2}$})
\prod_{\l=0}^{3}\lambda(\gam+\hbox{${1\over3}$},1-3\be_\l)
\ \lambda(-\gam-\hbox{${1\over3}$},\hbox{${3\over2}$}+3\be_\l)
$$
where we have defined $X\equiv[\mu\,{\Gamma({1\over3})/\Gamma({2\over3})}
]^{-(3\gam+1)/2}({2\over3})^{\gam-{2/3}}$,\
 $\gam\equiv\sum_{i=1}^3\be_i$ and $\be_0\equiv-{1\over6}$, as
well as
$$\eqalign{
\lambda(z,b)=(2\pi)^{z/2}&\exp[-3z+(\log3-E)f(z)]\ \prod_{\l=0}^\infty\left(
\vphantom{{({3\over2}z+b+3\l+k)\over(b+3\l+k)}}\right.
\exp\left[-3z+{f(z)\over \l+1}\right]\cr
&\left.\left[\prod_{k=0}^2
{({3\over2}z+b+3\l+k)\over(b+3\l+k)}{({3\over2}z+b+3\l+k+{3\over2})\over
(b+3\l+k+{3\over2})}\right]^{\l+1}\right)\ .\cr}
$$
Here $f(z)={3\over4}z^2+(b-{5\over4})z$, and $E=.577\ldots$
is Euler's constant.
The $\lam$-function in Eq.~(8) satisfies
$\lambda(z+1,b)=\Gamma(\hbox{$\ths$}z+b)\lambda(z,b)$, amongst other
properties.

One of the most interesting, and presumably most physically significant
features of this amplitude is the infinite
set of poles which it contains. Keeping in mind the restriction
$\be_i\ge-{1\over6}$, the $\lambda$-functions introduce two
sequences of poles of order
$\l+1$ at $\be_i=\l+{m\over3}+{1\over3},$ and $\be_i=\l+{m\over3}+{5\over6}$
for $\l=0,1,2,\ldots$ and $m=0,1,2$.
These are leg poles depending on the external momenta of the
individual tachyons. From the relation $k_i^2=2-\be_i-3\be_i^2$,
one finds that the poles in the first sequence with $m=0$, and all
of the poles in the second sequence, do {\elevenit not} correspond to the
physical mass shell of external string states.
The prefactor $\Gamma({1+3\gam\over2})\Gamma({1-3\gam\over2})$
also contains a sequence of simple poles at $\gam=-{1/3}$ and
${1/3}+2n$ with $n$, an non-negative integer.
In fact, the latter are cancelled by zeroes appearing in the
product of $\lam$-functions, in particular the $\l=0$ term.
There remains a single simple pole in the amplitude at
$\gam=-{1/3}$. In contrast to the previous leg poles, this pole
depends collectively on all of the external momenta.
Note that $\gam=-{1/3}$ is precisely the value of $\gam$ at which the
amplitude is independent of the cosmological constant, or equivalently
the world-sheet area.
This scale independence appears then to be the physical
reason for the pole, but its occurrence is an entirely stringy feature.

\vglue 0.6cm
\line{\elevenbf 3. Conclusions and prospects \hfil}
\vglue 0.4cm
It has been our aim here to show that the effort expended on
the study of non-critical strings has
important physical consequences in critical string theories.
Any future progress in
non-critical string physics, or in quantum Liouville theory, will
be of use in understanding off-shell critical string physics.
In particular, advances in the conformal
field theory treatment of the Liouville correlators are clearly
needed.
There are no conceptual barriers to the extension of our results to
supersymmetric strings, or to open string theories.

A striking feature of the
amplitudes is the presence of poles that do not correspond to
excitations in the matter sector (even if combined with the ghost sector).
They may indicate the presence of
excitations that are entirely stringy in nature.
In particular, a pole which depends collectively
on all of the external momenta, such as that at $\gam=-1/3$,
is entirely unknown in the amplitudes one obtains from a field theory.
In field theories, the off-shell character of the amplitude is a function
of individual external states.
It is difficult then to imagine how this $\gam$ dependence could be reproduced
in a string field theory. Thus our results may indicate that
some fundamentally new framework, other than string field theory,
will be required to extend our
understanding of critical string theory beyond the Polyakov path integral.
\vglue 0.2cm
It is a pleasure to
thank the organizers of the Strings '93 for their hospitality.
R.C.M. was supported by NSERC of Canada, and Fonds FCAR du Qu\'ebec.
V.P. was supported by D.O.E. grant DE-FG02-90ER40542.
\vglue 0.6cm
\line{\elevenbf References \hfil}
\def\sl{\elevenit}
\vglue 0.4cm
\item{\pol.} A.M. Polyakov, {\sl Phys. Lett.} {\bf 103B} (1981) 207,
211.
\item{\papers.} R.C. Myers and V. Periwal,
{\sl Phys. Rev. Lett.} {\bf 70} (1993) 2841.
\item{\david.} N.E. Mavromatos and J.L. Miramontes, {\sl Mod. Phys.
Lett.}
{\bf A4} (1989) 1847; E. D'Hoker and P.S. Kurzepa, {\sl Mod. Phys.
Lett.}
{\bf A5} (1990) 1411; E. D'Hoker, {\sl Mod. Phys. Lett.} {\bf A6} (1991)
745;
F. David,{ \sl Mod. Phys. Lett.} {\bf A3} (1988) 1651;
J. Distler and H. Kawai, {\sl Nucl. Phys.} {\bf B321} (1989) 509.

\bye